\documentclass[aps,pra,a4paper,aps,twocolumn,superscriptaddress,groupedaddress]{revtex4}
\usepackage[dvips]{graphicx}
\usepackage{ae}
\usepackage[T1]{fontenc}
\usepackage[ansinew]{inputenc}
\usepackage{amsmath}
\usepackage{amssymb}
\usepackage{graphicx}
\usepackage{caption}
\usepackage{color}
\usepackage{xcolor}
\usepackage[colorlinks]{hyperref}
\usepackage{lscape}
\usepackage{amsfonts}
\usepackage{amssymb}
\usepackage{makeidx}
\usepackage{braket}
\usepackage{bm}
\usepackage{bbm}
\usepackage{hyperref}
\usepackage{dsfont}
\usepackage{enumerate} 
\usepackage{float} 
\usepackage{mathrsfs}
\usepackage{mathtools,nccmath}
\usepackage{xfrac}

\begin{document}
\title{Bell's inequality violation by dynamical Casimir photons in a superconducting microwave circuit}

\author{Riddhi Chatterjee}
\email{riddhichatterjee27@gmail.com}
\affiliation{S. N. Bose National Centre for Basic Sciences, Block JD, Sector III, Salt Lake, Kolkata 700106, India.}

\author{A. S. Majumdar}
\email{archan@bose.res.in}
\affiliation{S. N. Bose National Centre for Basic Sciences, Block JD, Sector III, Salt Lake, Kolkata 700106, India.}

\begin{abstract}
We study the Bell's inequality violation by dynamical Casimir radiation with pseudospin measurement. We consider a circuit quantum electrodynamical set-up where a relativistically moving mirror is simulated by variable external magnetic flux in a SQUID terminating a superconducting-microwave waveguide. We analytically obtain  expectation values of the Bell operator optimized with respect to channel orientations, in terms of the system parameters.  We  consider the effects of local noise in the microwave field modes, asymmetry between the  field modes resulting from nonzero detuning, and  signal loss. 
Our analysis provides ranges of the above experimental parameters for which Bell violation can be observed. We show that Bell violation can be observed in this
set-up up to $40 mK$ temperature as well as up to $65$\% signal loss.
\end{abstract}

\maketitle

\section{Introduction}
Fluctuaing vacuum of quantum fields produces particles when the geometry of the background space-time is curved or changes with time. Quantized vacuum emits particles known as Hawking radiation \cite{hawk} due to gravity in the space-time of black holes. Cosmological particle production can be explained as the radiation by quantum vacuum in the expanding universe \cite{parker,birrel_davis,cosmo_particle}. Radiation by quantized vacuum  occurs inside cavities \cite{moore} and in  free space \cite{dewitt,fulling_davis1} in presence of moving mirrors or when the field is subjected to  time dependent boundary conditions. This phenomena is called the Dynamical Casimir effect (DCE) or Moore-DeWitt effect \cite{moore, dewitt}. 

The physics of these quantum radiations have profound significance in the foundations of quantum  theory in curved space-time \cite{birrel_davis}, which predicts that the emitted particles have blackbody distribution \cite{parker2,carlip,fulling_davis2}.  Thermodynamics of black holes \cite{carlip} may give rise to the information paradox which has been a
subject of much debate \cite{unruhwald}. This has motivated
the study  quantum correlations in Hawking radiation \cite{maldac,mahaj,raju}. Extensive studies have been performed on various foundational and information theoretic aspects in curved space-time \cite{maldac,mahaj,raju,alice_bh,fuentes_cosmo,celeri,lock_dce,schw,
sbfp,rc3}. However, experimental detection of such radiation in nature seem to be out of reach of current technology.

With the development of quantum material technology, it has been possible to create analogue DCE in the laboratory by varying bulk properties of materials such that time dependent boundary condition are induced (quantum simulation of moving mirror) in the medium containing the quantum field \cite{jo1,jo2,jo3,parn}. The time dependent boundary condition makes the vacuum state of the field evolve  into a superposition of states of various numbers of particles, as denoted mathematically by a Bogolyubov transformation \cite{fulling_davis1,fulling_davis2,scholar}. This results in emission of Casimir photon pairs. If the background space-time is flat apart from the boundary condition, the emission can be localized as the radiation from a moving mirror \cite{moore,fulling_davis1,fulling_davis2,scholar}. A relativistically moving mirror causes rapid nonadiabatic modulation of quantum field modes. To accommodate the modulated modes, the quantum vacuum throws up particles that follows Planck's distribution. 

Though for the case of non-relativistically moving mirror, particles can be created through nonuniform acceleration, the rate of particle production is extremely small \cite{moore}.
Moreover, a cavity set-up is generally introduced to
obtain parametric amplification. In practice, making a mechanical mirror move near the speed of light 
has been technologically challenging for a long time. DCE was first observed experimentally \cite{jo3,parn} in superconducting microwave circuits, by simulating the relativistic motion of mirrors \cite{jo1,jo2}. Here the superconducting transmission line is interrupted by superconducting quantum interference devices (SQUIDs). Inductance of the SQUID is ultra-rapidly tuned by making high frequency modulation of external magnetic flux threading the SQUID. Change in the inductance causes change in the electrical length of the transmission line. Hence, a time dependent boundary condition similar to that
induced by a  relativistically moving mirror is imposed to the quantized microwave field in the transmission line through the screening current flowing through the SQUID. In the above circuit quantum electrodynamical (cQED) set-up, the simulated velocity of the mirror can reach $\sim 10^6\ m\ s^{-1}$, and
the number of photons generated per second is $10^5$ \cite{jo2}.  DCE can also be implemented in the optomechanical set-up \cite{macri}.

DCE entangles the field modes and the nature of the radiation is nonclassical \cite{jo4}. Entanglement in the DCE is a consequence of the energy and momentum conservation, just like in the parametric down conversion process.  Quantum correlations generated through DCE can be used as a resource for quantum information processing. Even in a realistic cQED set-up where the background noise is present, the microwave radiation can be nonclassical \cite{jo4}. Apart from its significance as a resource for quantum tasks, the above experimental set-up provides a platform for simulation of fundamental phenomena \cite{sbfp,rc3,cosm_cqed}. Quantum features such as entanglement, coherence and discord of noisy Casimir photons have been studied with respect to various circuit parameters \cite{dcecoh,discord_dce}. Gaussian interferometric power, and steering have also been studied in presence of noise \cite{symst,asymst,asymst_dech}. 

Bell nonlocality is the strongest of all quantum correlations, manifesting violation of local realism \cite{bell1,bell2}. Bell presented his seminal work 
as a quantitative formulation of the ideas  of Einstein, Podolsky and Rosen (EPR) that suggest quantum mechanics as an incomplete theory \cite{epr}. EPR 
based their argument on states having  continuous spectrum in phase space. 
However, the formalisms of Bohm \cite{bohm} and  Bell \cite{bell1,bell2}  
were based on discrete dimensional systems, as  did the work of Clauser, Horne, Shimony, and Holt (CHSH) \cite{chsh}. Subsequently, Bell's inequality violation has been extensively investigated both in discrete and continuous variable systems \cite{aspect,reidbl1,qoptics_millburn,kimble,revzen1,jwpan,revzen2,
revzen3,dorantes_degenerate}. The significance of Bell nonlocality in the
security of  quantum cryptography \cite{cryp1,cryp2,cryp3,cryp4} has come under sharper 
focus with the development of device independent quantum key distribution 
protocols \cite{vazirani, miller, rene}.   Bell's inequality violation has  been studied in diverse domains ranging from  fundamental phenomena such as Unruh effect \cite{unruh_bell} and cosmic photons \cite{cosm_bell1,cosm_bell2} to
 the case of laboratory experiments using quantum materials \cite{bellmatter1,bellmatter2}.

In this work we investigate Bell's inequality violation by dynamical Casimir radiation using non-Gaussian pseudospin measurements. Our primary motivation arises
from the fact that the  study of Bell's inequality in context of particle production through  time dependent boundary conditions is hitherto unexplored in the literature, even though the system of circuit quantum electrodynamics with a moving mirror  can be  implemented in the laboratory \cite{jo3,parn}. Fundamentally, the dynamical space-time background generally produces Gaussian states. The quantum state of the microwave Casimir photons is a two mode squeezed thermal state \cite{jo1,jo2,jo3,jo3,jo4,dcecoh,discord_dce,symst,asymst,asymst_dech}, though a NOON state has also been engineered using DCE array \cite{noon}.
 Several studies have been performed on quantum correlations of Casimir photons based on Gaussian measurement \cite{jo4,dcecoh,discord_dce,symst,asymst,asymst_dech}. Here we explore  quantum correlations of DCE radiation with measurement beyond the Gaussian regime. Non-Gaussian measurement is a significant tool in quantum information such as quantum teleportation \cite{revzen_telep}, steering and cryptography \cite{pspin_nha,pspin_adesso,singh_bose}, non-Gaussian state preparation from Gaussian state \cite{treps_nong}. As it is possible to implement DCE in the laboratory,  our analysis should be relevant to understand the efficiency of DCE as a resource for quantum information.

Our approach is based on employing pseudospin mesurements represented in configuration space \cite{revzen2,revzen3}. Such measurements have been used to study Bell nonlocality of two mode squeezed vacuum \cite{revzen2,revzen3}, cosmic photons \cite{cosm_bell2} and quantum steering of two mode squeezed thermal state \cite{pspin_adesso}. The pseudospin operators represented in configuration space is  easier to handle  compared to its representation in the number basis \cite{pspin_adesso}. It may be noted that optimization of the expectation value of the Bell operator in configuration space involves additional parameters compared to that in the number state representation\cite{revzen2}. Pseudospin measurement in configuration space has also been used to study nonlocality of different classes of multimode Gaussian states \cite{pspin_mltnonloc}, enhancement of nonlocality \cite{pspin_enhnonloc}, and quantum teleportation \cite{revzen_telep}. Our aim is to use this measurement to explore Bell-CHSH inequality violation by Casimir photon pairs in the cQED set-up. We investigate how the optimal value of Bell violation depends on various circuit parameters. Specifically, we consider the effect of local noise, nonzero detuning and signal loss.

Our work is organized as follows: In section \ref{sec2} we  describe DCE in 
the cQED set-up. In \ref{sec3} we  study violation of the Bell-CHSH inequality by Casimir photon pairs generated via the cQED set-up described in section \ref{sec2}. We explore the dependence of optimal Bell violations on different system parameters. In section \ref{sec4} we  study the robustness of the Bell violation under signal loss. In section \ref{sec5} we present a summary
of our main results along with some concluding remarks.


\section{DCE in superconducting circuit}\label{sec2}

\subsection{System specification}
We consider the cQED setup described in \cite{jo1,jo2,jo3,jo4}. A superconducting coplanar waveguide (CPW) with characteristic
capacitance $C_0$ and inductance $L_0$ per unit length is terminated to ground (say at $x=0$) by a SQUID loop threaded by external magnetic flux $\phi_{ext}(t)$. The quantized microwave field inside the waveguide is described by its flux field  $\phi^i (x,t)$ that obeys the 1-D massless Klein-Gordon equation  \begin{equation}\label{kg}
\left(\partial_{xx} - v^{-2} \partial_{tt} \right) \phi^i (x,t) = 0,
\end{equation}
with Dirichlet boundary condition at $x=0$. In the framework of input-output theory, the total flux field in CPW transmission line is \begin{equation}\label{flux_tot}
\begin{aligned}
\phi (x,t) = \phi^{in} (x,t) + \phi^{out} (x,t)
\end{aligned}
\end{equation}
where $\phi^{in} (x,t)$ and $\phi^{out} (x,t)$ are the incoming (right moving) and outgoing (left moving) component of total flux field.\par
In terms of the second quantized solution of Klein-Gordon equations, the expression (\ref{flux_tot}) can be written as \cite{jo1,jo2,jo4,ugalde_thesis} 
\begin{equation}\label{kgsol2}
\begin{aligned}
\phi (x,t) = \sqrt{\mfrac{\hbar Z_0}{4\pi}} \int_{-\infty}^{\infty} \frac{d\omega}{\sqrt{|\omega|}} \Big[  \hat{a}_{\omega}^{in}\ e^{-i(\omega t - k_{\omega} x)} \\ +\  \hat{a}_{\omega}^{out}\ e^{-i(\omega t + k_{\omega} x)} \Big],\ (x<0).
\end{aligned}
\end{equation}
$\hat{a}_{\omega}^{in}$  and $\hat{a}_{\omega}^{out}$ are annihilation operators for modes of frequency $\omega$, propagating with velocity $v$ to the right (incoming) and left (outgoing) respectively. $\left[ \hat{a}_{\omega}^{in(out)}, (\hat{a}_{\omega'}^{in(out)})^{\dagger} \right] = \delta (\omega - \omega')$ and we  use the convention $\hat{a}_{- \omega} = \hat{a}_{\omega}^{\dagger}$. The velocity $v = \omega / k_{\omega} = 1/\sqrt{C_0 L_0} $ and $Z_0 = \sqrt{L_0 / C_0}$ is the characteristic impedance of CPW. \par
At large plasma frequency, when charging energy is much smaller than the external flux dependent effective Josephson energy $E_J(t) = E_J(\phi_{ext}(t))$, the SQUID is a passive element that provides the following boundary condition at $x=0$, to the flux field inside CPW line \cite{jo1,jo2} \begin{equation}\label{bd2}
\begin{aligned}
\phi (0,t) + L_{eff}(t)\ \partial_x \phi(x,t) \bigg\vert_{x = 0} = 0,
\end{aligned}
\end{equation}
where $L_{eff}(t) = L_J(t)/L_0$ and $L_J = \left(\mfrac{\phi_0}{2\pi}\right)^2 \mfrac{1}{E_J(t)}$ is the tunable Josephson inductance of the SQUID. $\phi_0 = (h/2e)$ is the magnetic flux quantum. The  boundary condition at $x=0$  depends only upon tunable Josephson inductance of the SQUID \cite{jo1,jo2,jo3}, which creates a mirror at an effective length $L_{eff}(t)$ from the physical end of the CPW line. For sinusoidal modulation $E_J(t) = E_J^0 (1 + \epsilon \sin{\omega_d t})$ with driving amplitude $\epsilon$ and driving frequency $\omega_d$ \cite{jo1,jo2},  the effective length modulation amplitude is $\delta L_{eff} = \epsilon L^0_{eff}$, where $L^0_{eff} = L_{eff}(0)$. So, the effective velocity of the mirror is $v_{eff} = \omega_d L^0_{eff}$. When $v_{eff}$ is a significant fraction of the velocity $v$ of light in CPW line, nonadiabatic modulation occurs in the field modes resulting in significant amount of photon pair production.\par
In case of weak harmonic drive (perturbative regime) \cite{jo1,jo2}, $\epsilon E_J^0 \ll E_J^0$ and the output photon-flux density has a parabolic spectrum with a maximum at $\omega_d/2$. Output photon pairs are correlated with frequency $\omega_{\pm}$, where $\omega_+ + \omega_- = \omega_d$. The simplest choice is $\omega_{\pm} = \omega_d/2 \pm \delta \omega$, where $ \delta \omega$ is the detuning parameter. Using equations (\ref{kgsol2}) and (\ref{bd2}) and applying scattering theory, the Bogolyubov transformation between incoming and outgoing modes can be obtained analytically in the perturbative regime \cite{jo2,jo4,ljr} \begin{equation}\label{bog1}
\begin{aligned}
\hat{a}_{\omega_{\pm}}^{out} = - \hat{a}_{\omega_{\pm}}^{in} - i f (\hat{a}_{\omega_{\mp}}^{in})^{\dagger}
\end{aligned}
\end{equation}
where \begin{equation}\label{bog2}
\begin{aligned}
f = \frac{\epsilon L^0_{eff}\ \sqrt{\omega_+ \omega_-}}{v}.
\end{aligned}
\end{equation}
Pair production results in two mode squeezing of the output field \cite{jo2,jo4}. So, if the input state $\phi^{in}$ in equation (\ref{flux_tot}) is a vacuum state, the output DCE state $\phi^{out}$ is ideally a two mode squeezed vacuum.

\subsection{Covariance matrix of input/output modes}
Let us consider the DCE input/output states in the framework of Gaussian covariance matrix formalism  \cite{jo4}. In our work we follow the convention of \cite{adesso_review}.  Input/output state can be written in terms of covariance matrix (CM) \begin{equation}
\begin{aligned}
V_{\alpha \beta} = \langle \hat{R}_{\alpha}\hat{R}_{\beta} + \hat{R}_{\beta}\hat{R}_{\alpha} \rangle - 2 \langle \hat{R}_{\alpha} \rangle \langle \hat{R}_{\beta} \rangle ,
\end{aligned}
\end{equation}
where $\hat{R}=(\hat{q}_-,\hat{p}_-, \hat{q}_+,\hat{p}_+)^T$ is a vector containing field quadrature elements with $[\hat{q}_{\alpha},\hat{p}_{\beta}]=i\delta_{\alpha \beta}$ and \begin{equation}\label{in_quad}
\begin{aligned}
\hat{q}_{\pm}^{in (out)} &= \frac{\hat{a}_{\omega_{\pm}}^{in (out)} + (\hat{a}_{\omega_{\pm}}^{in (out)})^{\dagger}}{\sqrt{2}} \\ 
\hat{p}_{\pm}^{in (out)} &= - i\  \frac{\hat{a}_{\omega_{\pm}}^{in (out)} - (\hat{a}_{\omega_{\pm}}^{in (out)})^{\dagger}}{\sqrt{2}}.
\end{aligned}
\end{equation}
where we have restricted ourself to a pair of entangled modes $\lbrace \pm \rbrace$ with frequencies adding up to the driving frequency. Nonclassicality of DCE radiation originates from the entanglement of Casimir photon pairs \cite{jo4}. Ideal input state will be a vacuum state which is impossible to create in practical situation. So, we will use a quasi-vacuum state, containing small number of thermal photons $\Big\lbrace n^{th}_{\pm} = \Big( e^{\tfrac{\hbar \omega_{\pm}}{kT}} - 1 \Big)^{-1} \Big\rbrace$ \cite{jo4,discord_dce}, as the input state. 

Note that the choice of
the initial thermal state modifies the Green's function and the power spectrum,  and the frequencies of incoming and outgoing signal are doppler shifted
\cite{dce_dchneu}.
 However, the effect on the correlation of the outgoing modes is neglibible
 for our chosen temperature range \cite{parn}. 
   Hence, correlation of the output modes are observed in experiments despite  the fact that the effects of temperature are not explicitly monitored during the experiments \cite{dce_dchneu, parn}. 
The quadrature elements have zero $1^{st}$-moment. CM of input field is given by \begin{equation}\label{in_cov}
\begin{aligned}
V_{in} = \begin{pmatrix}
n_0 & 0 & 0 & 0 \\
0 & n_0 & 0 & 0 \\
0 & 0 & m_0 & 0 \\
0 & 0 & 0 & m_0
\end{pmatrix}
\end{aligned}
\end{equation}
where \begin{equation}\label{in_cov1}
\begin{aligned}
n_0 &= (2 n^{th}_- + 1) \\
m_0 &= (2 n^{th}_+ + 1).
\end{aligned}
\end{equation}
Using equations (\ref{bog1},\ref{bog2},\ref{in_quad},\ref{in_cov},\ref{in_cov1}), CM of output field is obtained (in standard form) as \begin{equation}\label{out_cov}
\begin{aligned}
V_{out} = \begin{pmatrix}
n & 0 & r & 0 \\
0 & n & 0 & -r \\
r & 0 & m & 0 \\
0 & -r & 0 & m
\end{pmatrix}
\end{aligned}
\end{equation}
where \begin{equation}\label{out_cov1}
\begin{aligned}
n &= (2 n^{th}_- + 1) + f^2 (2 n^{th}_+ + 1) \\
m &= (2 n^{th}_+ + 1) + f^2 (2 n^{th}_- + 1) \\
r &= 2 f (n^{th}_+ + n^{th}_- + 1).
\end{aligned}
\end{equation}
$V_{out}$ corresponds to a two mode squeezed thermal state with squeezing parameter $2f$.

\section{Bell violation by DCE radiation}\label{sec3}
Wel study Bell's inequality violation by the output DCE radiation described by the CM in equations (\ref{out_cov},\ref{out_cov1}), using the definition of pseudospin measurement represented in configuration space \cite{revzen2,revzen3} \begin{equation}
\begin{aligned}
\hat{\Pi}_x &=& \int^{\infty}_0 dq \left[ \ket{\mathcal{Q}^+}\bra{\mathcal{Q}^-} + \ket{\mathcal{Q}^-}\bra{\mathcal{Q}^+} \right] \\
\hat{\Pi}_y &=& i \int^{\infty}_0 dq \left[ \ket{\mathcal{Q}^+}\bra{\mathcal{Q}^-} - \ket{\mathcal{Q}^-}\bra{\mathcal{Q}^+} \right] \\
\hat{\Pi}_z &=& -\int^{\infty}_0 dq \left[ \ket{\mathcal{Q}^+}\bra{\mathcal{Q}^+} - \ket{\mathcal{Q}^-}\bra{\mathcal{Q}^-} \right]
\end{aligned}
\end{equation}
where the channels $\ket{\mathcal{Q}^+}$ and $\ket{\mathcal{Q}^-}$ are given by
\begin{equation}
\begin{aligned}
\ket{\mathcal{Q}^+} &=& \frac{1}{2} \left[ \ket{q} + \ket{-q} \right] \\
\ket{\mathcal{Q}^-} &=& \frac{1}{2} \left[ \ket{q} - \ket{-q} \right].
\end{aligned}
\end{equation}
$\lbrace \hat{\Pi}_x,\hat{\Pi}_y,\hat{\Pi}_z \rbrace$ satisfy $SU(2)$ algebra. Following \cite{revzen2}, the Bell operator is defined as \begin{equation}
\begin{aligned}
\mathcal{B} = \vec{a} \cdot \hat{\Pi}^{(-)} \otimes \vec{b} \cdot \hat{\Pi}^{(+)} + \vec{a} \cdot \hat{\Pi}^{(-)} \otimes \vec{b'} \cdot \hat{\Pi}^{(+)} + \\  \vec{a'} \cdot \hat{\Pi}^{(-)} \otimes \vec{b} \cdot \hat{\Pi}^{(+)} - \vec{a'} \cdot \hat{\Pi}^{(-)} \otimes \vec{b'} \cdot \hat{\Pi}^{(+)}
\end{aligned}
\end{equation}
where $\vec{a}, \vec{a'}, \vec{b}, \vec{b'}$ are the unit vectors that specify the orientation of the first and second channel respectively. $\hat{\Pi}^{(\pm)}$ designates channel $\hat{\Pi}$ applied on $\lbrace \pm \rbrace$ mode.\par
In order to calculate optimal Bell violation, first we perform orientational optimization of measurement directions following \cite{jwpan,revzen2,dorantes_degenerate}. We choose $\vec{a}, \vec{a'}, \vec{b}, \vec{b'}$ in spherical polar coordinate as \begin{equation}
\begin{aligned}
\phi_a = \phi_{a'} = \phi_b = \phi_{b'} = 0, \\
\theta_a = 0, \quad \theta_{a'} = \pi/2, \quad \theta_b = - \theta_{b'}.
\end{aligned}
\end{equation}
So, the Bell operator reduces to \begin{equation}
\begin{aligned}
\mathcal{B} = 2 \left( cos{\theta_b }\  \hat{\Pi}^{(-)}_x \otimes  \hat{\Pi}^{(+)}_x + sin{\theta_b}\ \hat{\Pi}^{(-)}_z \otimes  \hat{\Pi}^{(+)}_z \right).
\end{aligned}
\end{equation}
Maximizing over $\theta_b$ we get the maximal expectation value of $\mathcal{B}$, \begin{equation}\label{bellvalue}
\begin{aligned}
B_{max} = 2 \sqrt{ \left\langle \hat{\Pi}^{(-)}_x \otimes  \hat{\Pi}^{(+)}_x \right\rangle^2 + \left\langle \hat{\Pi}^{(-)}_z \otimes  \hat{\Pi}^{(+)}_z \right\rangle^2 }
\end{aligned}
\end{equation}
where $\langle \cdot \rangle$ is the expectation value of an operator for a given state. Bell violation occurs when \begin{equation}
B_{max} >2.
\end{equation}\par
We now calculate $B_{max}$ for our output state described by the covariance matrix $V_{out}$ in equations (\ref{out_cov},\ref{out_cov1}) and study how the value of $B_{max}$ depends upon various system parameters. In order to calculate expectation value of two mode pseudospin operators we  use the definitions \cite{revzen3,pspin_adesso} \begin{equation}
\begin{aligned}
\left\langle \hat{\Pi}^{(-)}_i \otimes  \hat{\Pi}^{(+)}_j \right\rangle = \frac{1}{(2\pi)^2} \int d^4X\ W_{out}(X) \cdot \\ W_{\hat{\Pi}^{(-)}_i}(q_-,p_-)\ W_{\hat{\Pi}^{(+)}_j}(q_+,p_+)
\end{aligned}
\end{equation}
where $X = (q_-, p_-, q_+, p_+)^T$, \begin{equation}
\begin{aligned}
W_{\hat{\Pi}_x}(q,p) = sgn{(q)} \\
W_{\hat{\Pi}_z}(q,p) = -\pi \delta(q) \delta(p)
\end{aligned}
\end{equation}
and the Wigner function of the output state is \begin{equation}
\begin{aligned}
W_{out}(X) = \frac{1}{\pi^2} \frac{1}{det(V_{out})} e^{- \left(X^T\ V^{-1}_{out}\ X\right)}.
\end{aligned}
\end{equation}
Plugging everything in equation (\ref{bellvalue}) we evaluate $B_{max}$ for output DCE radiation in terms of circuit parameters \begin{equation}
\begin{aligned}
B_{max} = 2 \Bigg(\frac{1}{\left(f^2-1\right)^4 \left(2 n_-+1\right){}^2 \left(2 n_++1\right){}^2} + \\ \frac{4 \tan ^{-1}\left(\frac{4 f^2 \left(n_-+n_++1\right){}^2}{\left(f^2-1\right)^2 \left(2 n_-+1\right) \left(2 n_++1\right)}\right){}^2}{\pi ^2}\Bigg)^{1/2}
\end{aligned}
\end{equation}
with $\lbrace n_{\pm}\rbrace$ defined above equation (\ref{in_cov}) and $f$ is given by equation (\ref{bog2}), as a function of driving amplitude $\epsilon$. Now we plot the variation of $B_{max}$ with respect to different experimental parameters in order to observe the Bell's inequality violation.

\begin{figure}[!htb]
\centering
\captionsetup{justification=raggedright,
singlelinecheck=false
}
\includegraphics[scale=.55]{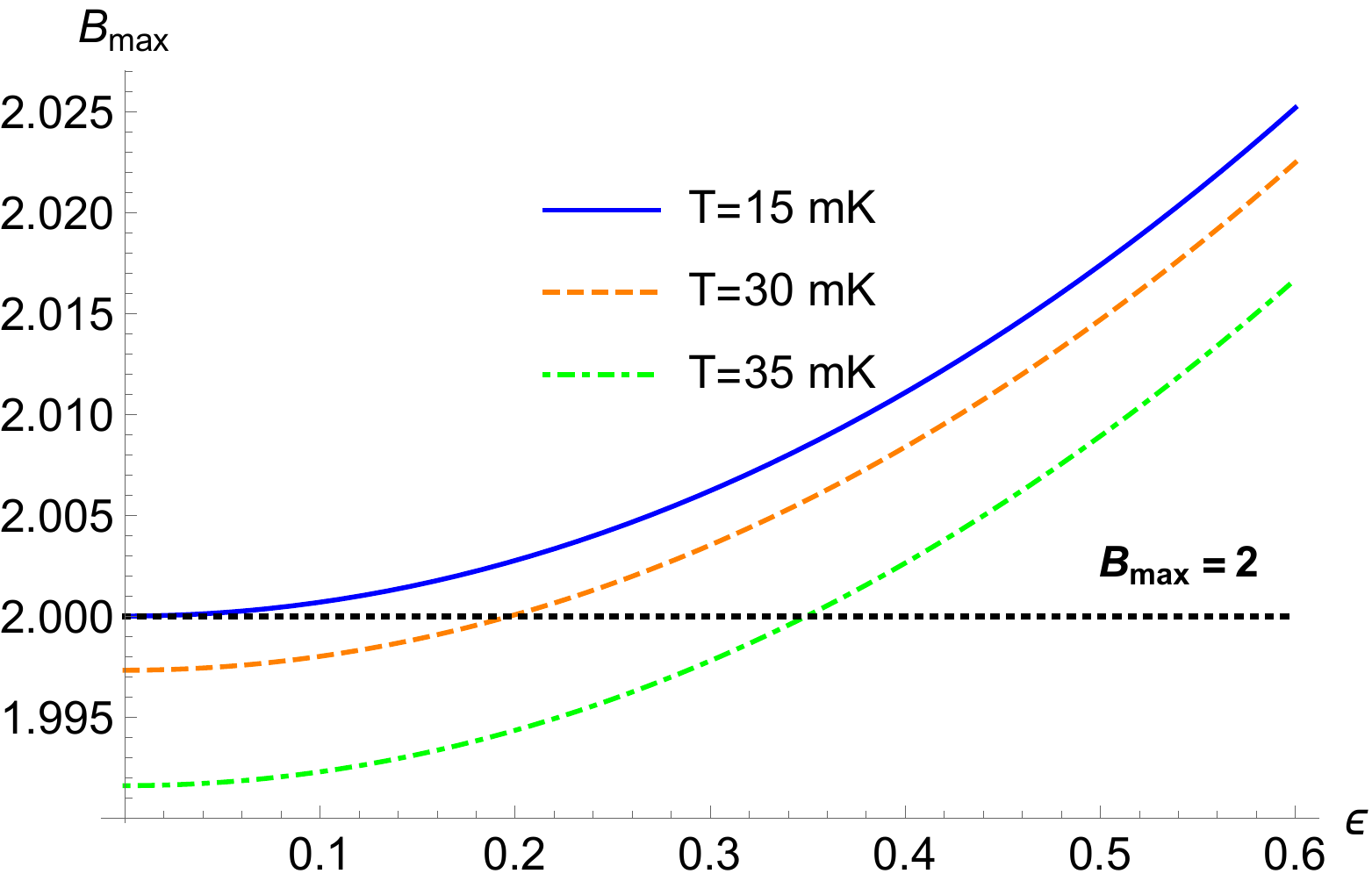}
\caption{\small{Variation of $B_{max}$ with the driving amplitude $\epsilon$ in different temperatures of the system. Driving frequency $\omega_d = 20\pi$ GHz. $v= 1.2 \times 10^8$ ms$^{-1}$. $L^0_{eff} = 0.5$ mm. Detuning $\delta \omega = 0$}.}
\label{1Bvepsilondet0T}
\end{figure}

Figure (\ref{1Bvepsilondet0T}) shows the variation of $B_{max}$ with increasing driving amplitude $\epsilon$ in different temperatures of the system. Here detuning is zero which means $n_- = n_+$ and hence both modes are symmetric and has equal local noises. The driving amplitude is considered upto $\epsilon = 0.6$, that corresponds to $f = 0.0786$ which is well inside perturbative regime. The plot shows that the value of $B_{max}$ has dropped significantly at $T = 35$ mK compared to its values at $T =15$ mK and $30$ mK. Also at $35$ mK, it requires significantly larger driving amplitude $> 0.35$, in order to observe Bell's inequality violation compared to the other two temperatures. The highest value of Bell violation obtained here, at $15$ mK and with $\epsilon = 0.6$, is 2.025.

\begin{figure}[!htb]
\centering
\captionsetup{justification=raggedright,
singlelinecheck=false
}
\includegraphics[scale=.5]{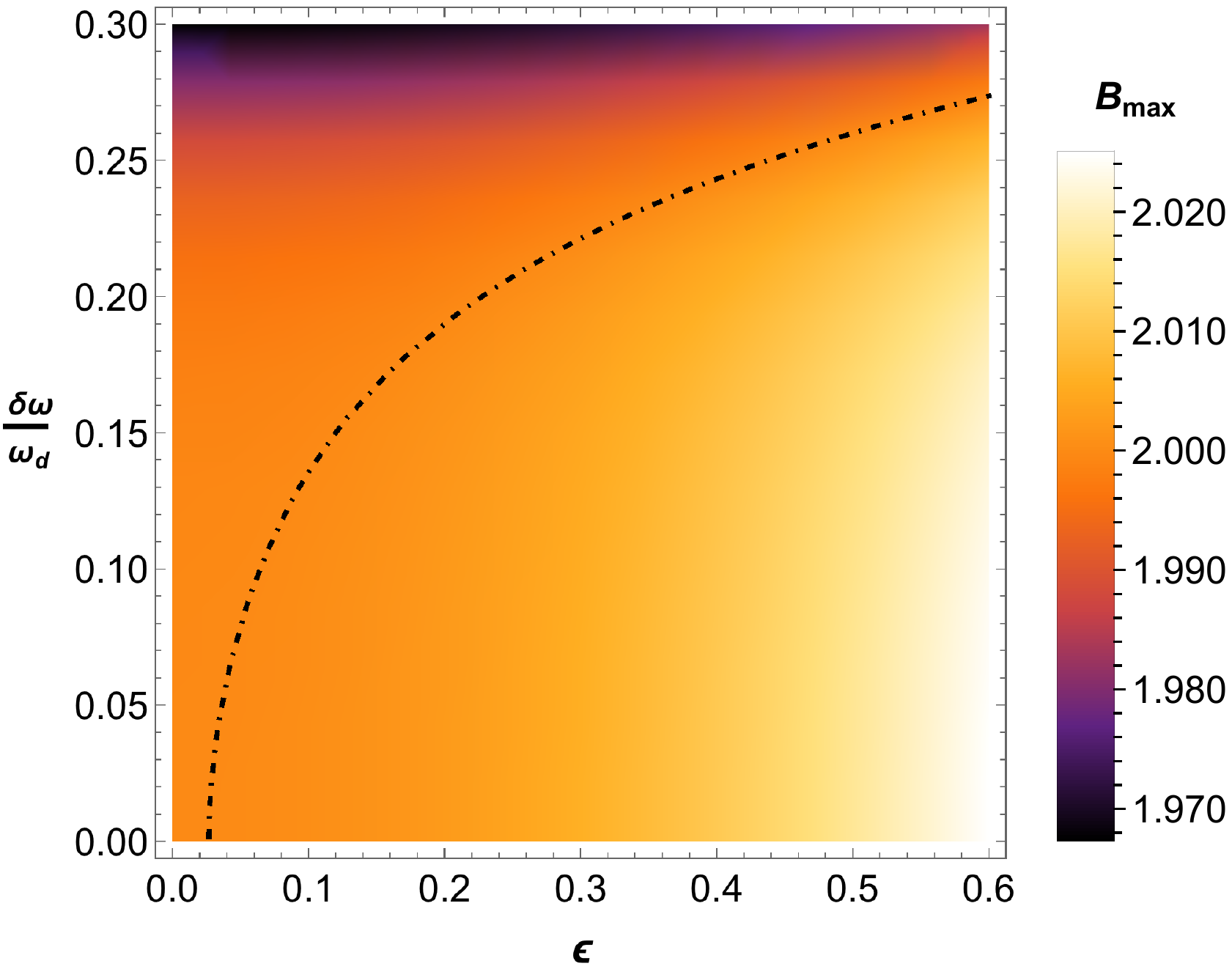}
\caption{\small{Variation of $B_{max}$ with respect to the driving amplitude $\epsilon$ and detuning $\delta \omega$ as the fraction of Driving frequency $\omega_d = 20\pi$ GHz. $v= 1.2 \times 10^8$ ms$^{-1}$. $L^0_{eff} = 0.5$ mm. $T = 20$ mK}.}
\label{2Bvepsilondetnz}
\end{figure}

Figure (\ref{2Bvepsilondetnz}) shows the variation of $B_{max}$ with increase in driving amplitude $\epsilon$ and $\frac{\delta \omega}{\omega_d}$ which is detuning expressed as a fraction of driving frequency $\omega_d$. The dash-dotted curve represents the points $B_{max} = 2$. So, the region on the right side of this curve violates Bell's inequality. The increase in detuning increases asymmetry between the two modes, decreasing the value of $B_{max}$. The plot also shows that in order to observe Bell's inequality violation in the chosen parameter range, detuning needs to satisfy $\frac{\delta \omega}{\omega_d} < 0.27$.

\begin{figure}[!htb]
\centering
\captionsetup{justification=raggedright,
singlelinecheck=false
}
\includegraphics[scale=.5]{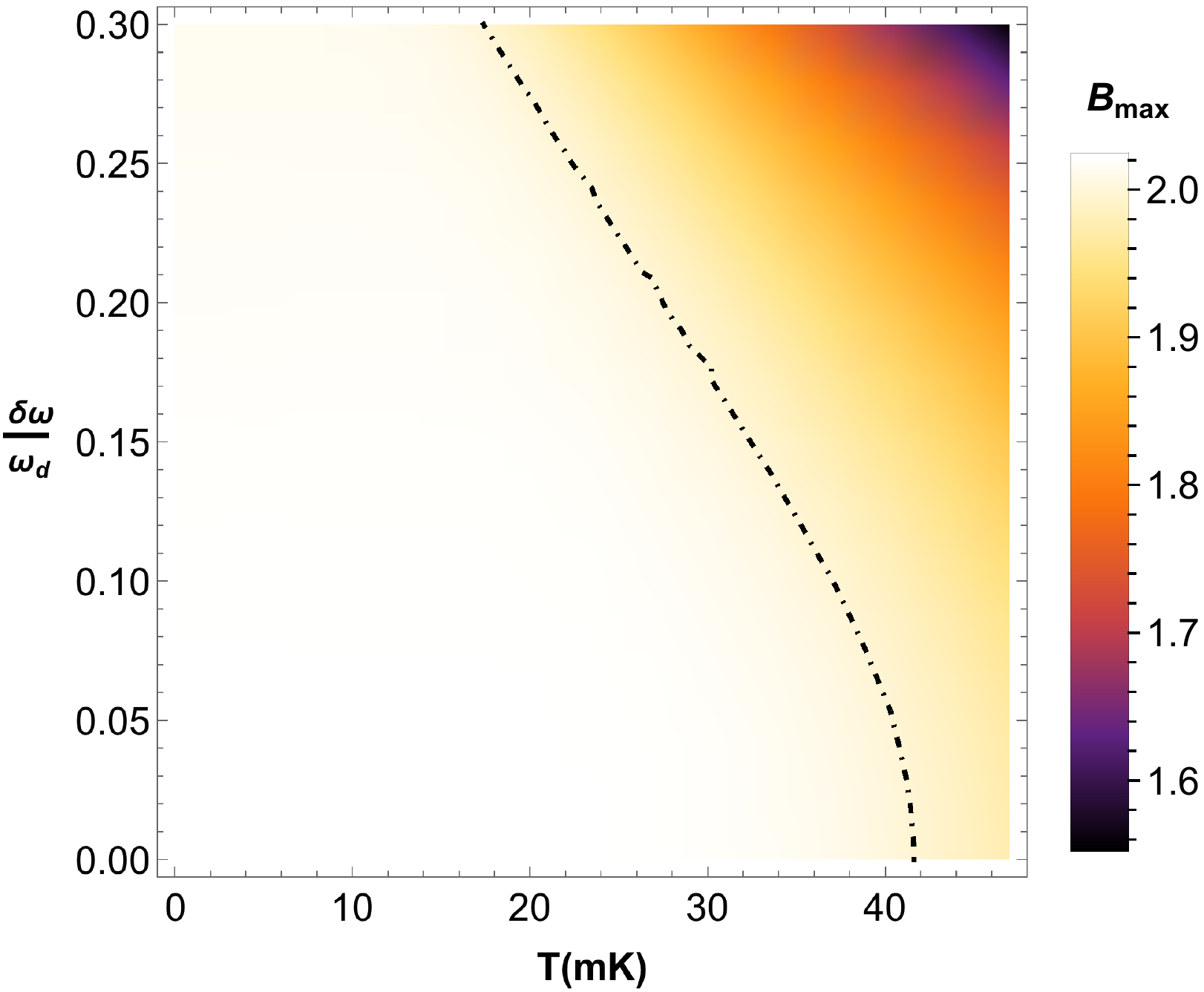}
\caption{\small{Variation of $B_{max}$ with respect to the temperature $T$ (mK) and detuning $\delta \omega$ as the fraction of driving frequency $\omega_d = 20\pi$ GHz, with $v= 1.2 \times 10^8$ ms$^{-1}$,  $L^0_{eff} = 0.5$ mm, and $\epsilon = 0.6$}.}
\label{3BvTdetnz}
\end{figure}

Figure (\ref{3BvTdetnz}) shows the variation of $B_{max}$ with increase in temperature $T$ and $\frac{\delta \omega}{\omega_d}$ which is detuning expressed as a fraction of driving frequency $\omega_d$. The dash-dotted curve represents the points $B_{max} = 2$. The region on the left side of this curve violates Bell's inequality. The plot indicates that at very low temperature, the effect of detuning on the value of $B_{max}$ is not significant and Bell violation is always achieved. This is because at low temperature, local noise is very small and hence asymmetry between the modes is very small even with significant value of detuning and driving amplitude $\epsilon$ (driving parameter $f$ is function of detuning and driving amplitude, see equation (\ref{bog2})). For temperature $20 - 30$ mK, the value of $B_{max}$  falls below $2$ for significant detuning. Around temperature $34$ mK, Bell violation is absent when $\frac{\delta \omega}{\omega_d}$ is approximately greater than
  $0.15$. For temperature $T \geqslant 42$ mK, Bell's inequality violation is completely absent for our chosen parameter range.

\section{Robustness under signal loss}\label{sec4}

In realistic scenarios, an experiment may suffer from imperfections. 
The two most relevant types of noise in the present set-up are
noise due to presence of thermal photons in the signal and signal loss in the transmission line.  In our analysis we have considered both the above types of noise.  The thermal noise that is observed to be present in DCE experiments implemented so far \cite{jo3,parn,broadband_dce}, is taken into account in our analysis. We further study tolerance of nonlocality of DCE radiation under another source of experimental defect,
{\it viz.}, photon loss which  is, in general, one of the most  studied defects in context of Bell violation \cite{BV_loss}.   Signal loss can occur  due to various reasons
 such as presence of impurities, measurement inefficiency, etc., \cite{mzi_cqed1},  and has been discussed in context of generation and measurement of DCE radiation \cite{jo3,parn,broadband_dce}. In our study we mimick the signal loss  by beam splitter operation and study the tolerance of Bell nonlocality of DCE radiation against such noise.
 
Our goal here is to  study if we can observe Bell's inequality violation in the experimental set-up under consideration, in presence of signal loss in one of the modes (say the first mode). We  apply pure loss channel on mode $\lbrace - \rbrace$ and obtain the output covariance matrix following the method of \cite{pspin_adesso}. We couple the state described by $V_{out}$, with a single mode vacuum (ancilla). Thus, the resultant covariance matrix is given by $V' = V_{anc} \oplus V_{out}$, where $V_{anc} = \mathds{1}_{2 \times 2}$ is the covariance matrix of the ancilla, with $\mathds{1}_{2 \times 2}$ being the $2 \times 2$ identity matix. We next transform the ancillary mode and $\lbrace - \rbrace$ mode though the beam splitter channel $B_s$.  We apply $B_s \oplus \mathds{1}_{2 \times 2}$ on $V'$ where 
\begin{equation}
\begin{aligned}
B_s = \left(
\begin{array}{cc}
 \sqrt{\eta }\ \mathds{1}_{2 \times 2} & -\sqrt{1-\eta }\ \mathds{1}_{2 \times 2} \\ [0.2cm]
 \sqrt{1-\eta }\ \mathds{1}_{2 \times 2} & \sqrt{\eta }\ \mathds{1}_{2 \times 2} \\
\end{array}
\right)
\end{aligned}
\end{equation}
with $\eta \in (0,1)$ being the transmission efficiency. Tracing out the ancillary modes leads us to obtain the covariance matrix of the output DCE radiation with signal loss on mode $\lbrace - \rbrace$, given by  \begin{equation}\label{loss_cov}
\begin{aligned}
V^L_{out} = \begin{pmatrix}
n' & 0 & r' & 0 \\
0 & n' & 0 & -r' \\
r' & 0 & m' & 0 \\
0 & -r' & 0 & m'
\end{pmatrix}
\end{aligned}
\end{equation}
where \begin{equation}\label{loss_cov1}
\begin{aligned}
n' &= \eta  \left(f^2 \left(2 n_++1\right)+2 n_-+1\right)-\eta +1 \\
m' &= (2 n^{th}_+ + 1) + f^2 (2 n^{th}_- + 1) \\
r' &= 2 f \sqrt{\eta } \left(n_-+n_++1\right)
\end{aligned}
\end{equation}
and $\lbrace n_{\pm} \rbrace$, $f$ have the same definitions as  in equation (\ref{out_cov1}). Following a similar procedure as in section \ref{sec3}, we find the expectation value of the Bell operator, optimized with respect to  the measurement orientations for the state described by the covariance matrix $V^L_{out}$ in terms of system parameters in presence of signal loss, given by \begin{equation}
\begin{aligned}
B^L_{max} = 2\sqrt{\frac{4}{\pi^2}\tan ^{-1}\left( A_1 / A_2 \right){}^2 + 1 /( A_2)^2}
\end{aligned}
\end{equation}
where \begin{equation}
\begin{aligned}
A_1 = 4 f^2 \eta  \left(n_-+n_++1\right){}^2 \\
A_2 = -4 f^2 \eta  \left(n_-+n_++1\right){}^2 +\\ \big( \left(2 f^2 n_-+f^2+2 n_++1\right)\times \\ \left(f^2 \eta +2 f^2 \eta  n_++2 \eta  n_-+1\right)\big)
\end{aligned}
\end{equation}
We  plot the variation of $B^L_{max}$ with respect to various system parameters, in order to observe the violation of Bell's inequality under signal loss.

\begin{figure}[!htb]
\centering
\captionsetup{justification=raggedright,
singlelinecheck=false
}
\includegraphics[scale=.5]{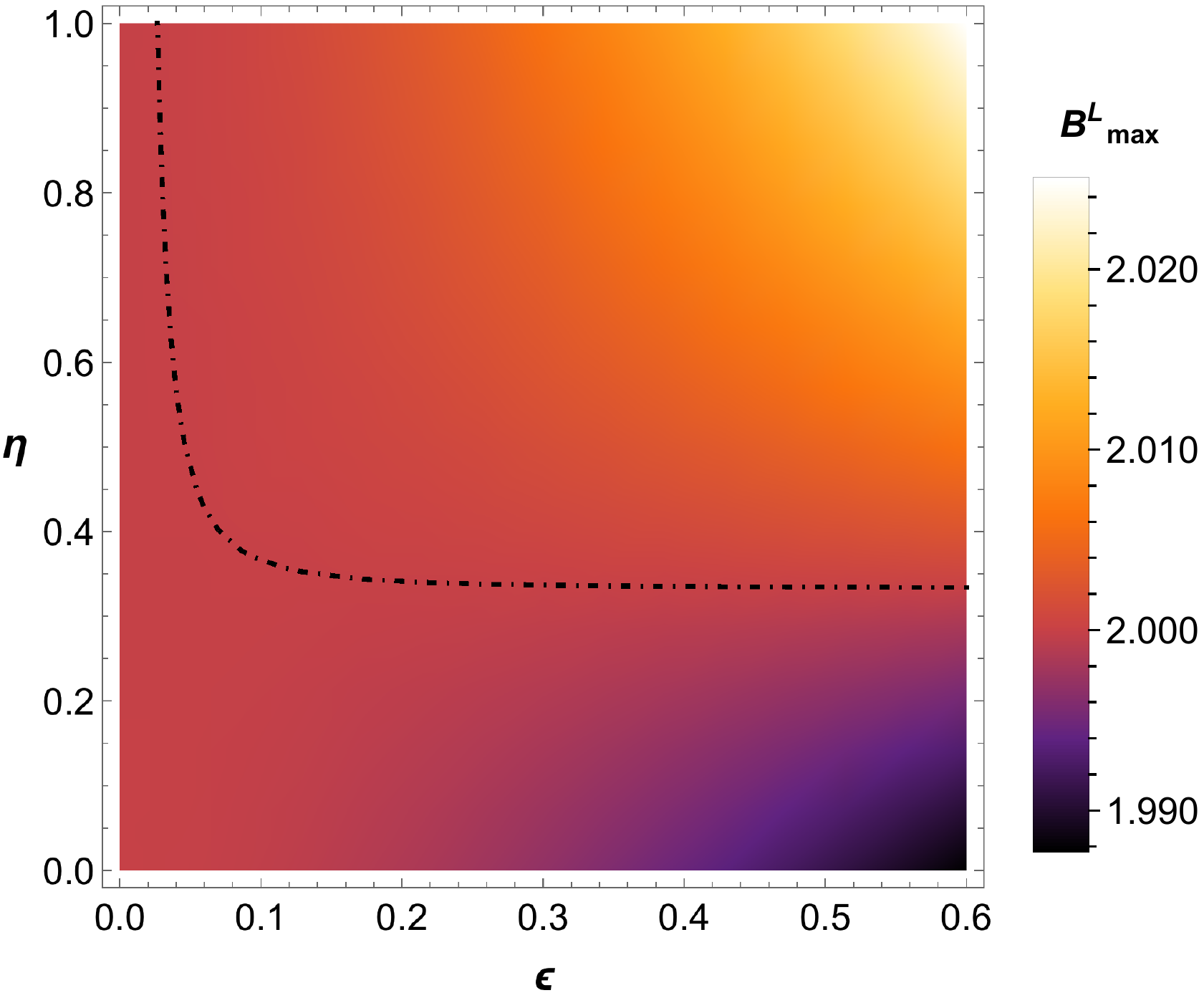}
\caption{\small{Variation of $B^L_{max}$ with respect to the driving amplitude $\epsilon$ and transmission efficiencys $\eta$. Driving frequency $\omega_d = 20\pi$ GHz. Detuning $\delta \omega = 0$. $v= 1.2 \times 10^8$ ms$^{-1}$. $L^0_{eff} = 0.5$ mm. $T = 20$ mK}.}
\label{4Bvepsilondet0L}
\end{figure}

Figure (\ref{4Bvepsilondet0L}) shows the variation $B^L_{max}$ with respect to the driving amplitude $\epsilon$ and transmission efficiency $\eta$ in absence of detuning. The dash-dotted curve represents the points $B_{max} = 2$. The region above this curve violates Bell's inequality. Below $\eta = 0.4$, the
threshold of Bell violation becomes less sensitive to the    increase of the driving amplitude. Bell violation is completely absent when $\eta < 0.35$, i.e., when the signal loss is greater than $65 \%$. The plot indicates that in order to observe noticeable Bell violation for our chosen range of parameters, the transmission efficiency should be greater than $0.4$.

\begin{figure}[!htb]
\centering
\captionsetup{justification=raggedright,
singlelinecheck=false
}
\includegraphics[scale=.5]{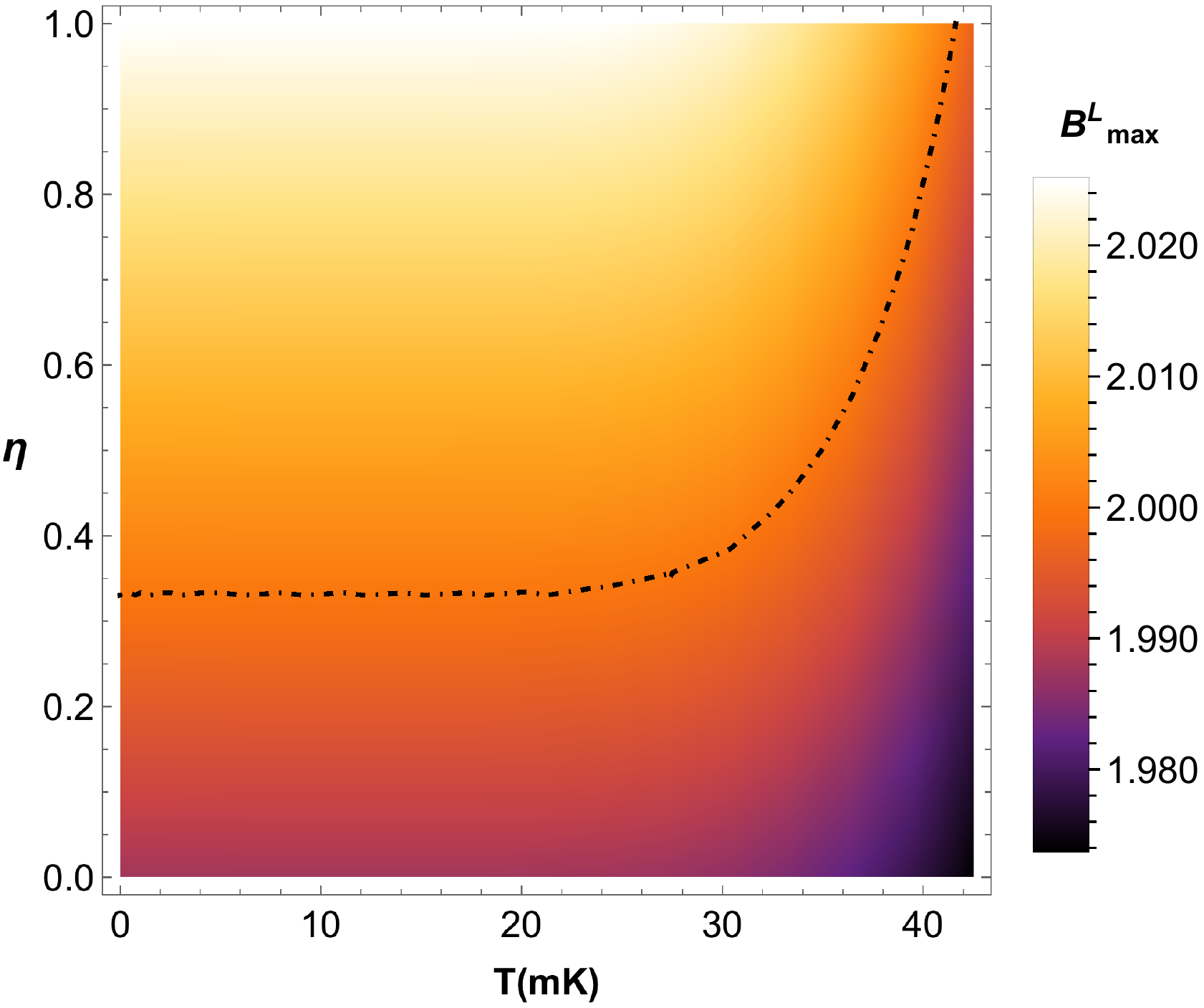}
\caption{\small{Variation of $B^L_{max}$ with respect to the temperature $T$ and transmission efficiency $\eta$. Driving frequency $\omega_d = 20\pi$ GHz. Detuning $\delta \omega = 0$. $v= 1.2 \times 10^8$ ms$^{-1}$. $L^0_{eff} = 0.5$ mm. $\epsilon = 0.6$}.}
\label{5BvTdet0L}
\end{figure}

Figure (\ref{5BvTdet0L}) shows the variation $B^L_{max}$ with respect to the temperature $T$ and transmission efficiency $\eta$ in absence of detuning. The dash-dotted curve represents the points $B_{max} = 2$. The region above this curve violates Bell's inequality. For temperature above $24$ mK, Bell violation is absent when the transmission efficiency $\eta < 0.35$ (signal loss $> 65 \%$ ). For higher temperatures, greater amount of transmission efficiency is needed in order to observe Bell violation. At temperatures near $40$ mK, the transmission efficiency has to be greater than $0.8$ (signal loss is required to be less than $20 \%$) for our chosen range of parameters.

\begin{figure}[!htb]
\centering
\captionsetup{justification=raggedright,
singlelinecheck=false
}
\includegraphics[scale=.5]{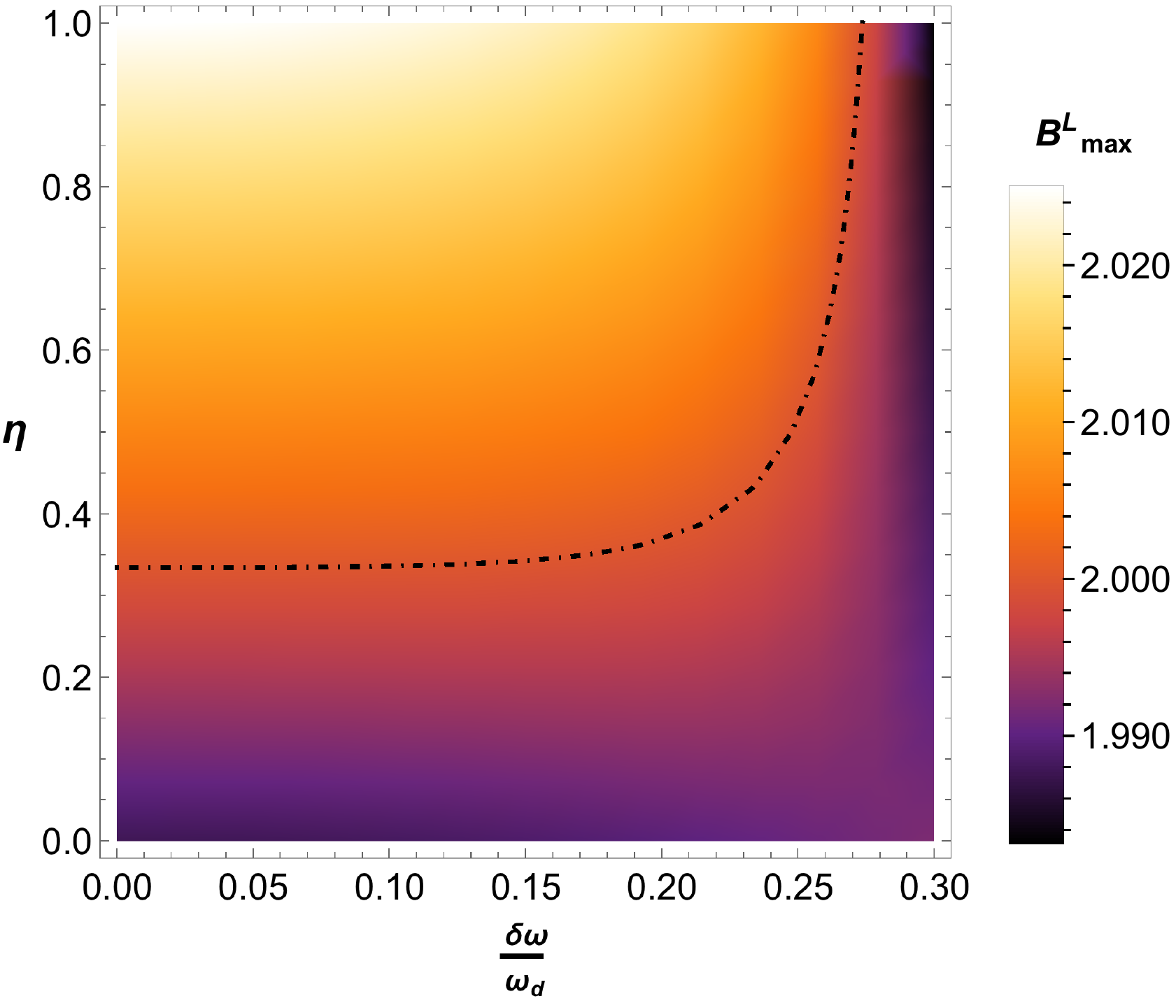}
\caption{\small{Variation of $B^L_{max}$ with respect to the transmission efficiency $\eta$ and the detuning as a fraction of the driving frequency $\omega_d = 20\pi$ GHz. $v= 1.2 \times 10^8$ ms$^{-1}$. $L^0_{eff} = 0.5$ mm. $\epsilon = 0.6$, $T = 20$ mK}.}
\label{6BvdetnzL}
\end{figure}

Figure (\ref{6BvdetnzL}) shows the variation of $B^L_{max}$ with respect to the transmission efficiency $\eta$ and the detuning as a fraction of the driving frequency $\omega_d$. The dash-dotted curve represents the points $B_{max} = 2$. The region above this curve violates Bell's inequality. The plot indicates that Bell violation occurs for $\eta \geq 0.35$ (signal loss $\leq 65 \%$ ) up to the detuning $\frac{\delta \omega}{\omega_d} = 0.2$. However for value of detuning $0.2 < \frac{\delta \omega}{\omega_d} < 0.27$, higher transmission efficiency is required. For detuning $\frac{\delta \omega}{\omega_d} > 0.27$ Bell violation is completely absent irrespective of the value of transmission efficiency. This is consistent with the result of  Figure (\ref{2Bvepsilondetnz}).

\begin{figure}[!htb]
\centering
\captionsetup{justification=raggedright,
singlelinecheck=false
}
\includegraphics[scale=.5]{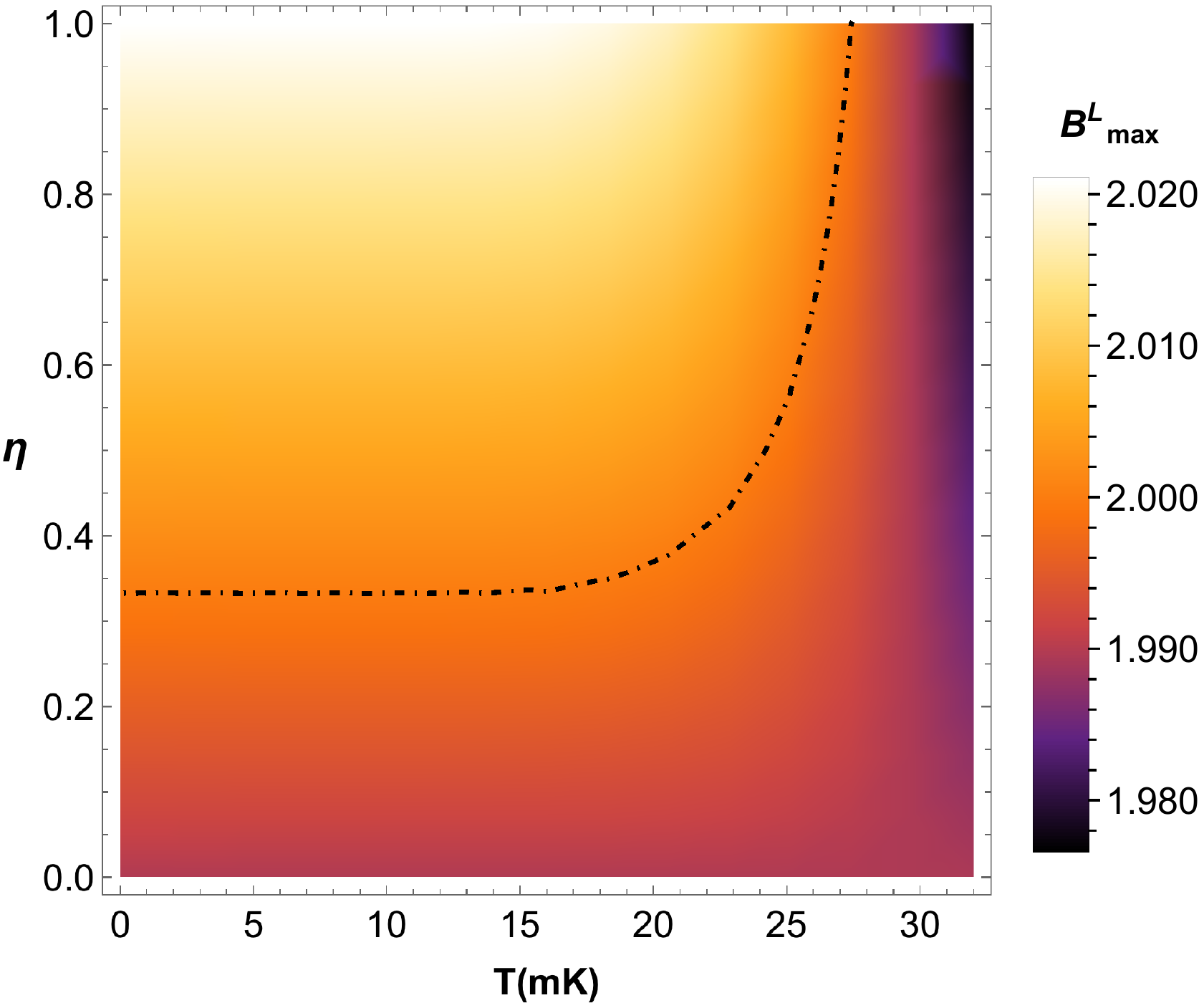}
\caption{\small{Variation of $B^L_{max}$ with respect to the temperature $T$ and transmission efficiency $\eta$. $(\delta \omega/\omega_d) = 0.2$ where $\delta \omega$ is the detuning and $\omega_d = 20\pi$ GHz is the driving frequency. $v= 1.2 \times 10^8$ ms$^{-1}$. $L^0_{eff} = 0.5$ mm. $\epsilon = 0.6$}.}
\label{7BvTdetnzL}
\end{figure}

Figure (\ref{7BvTdetnzL}) shows the variation of $B^L_{max}$ with respect to the temperature $T$ and the transmission efficiency $\eta$ in presence of detuning $\frac{\delta \omega}{\omega_d} = 0.2$. The dash-dotted curve represents the points $B_{max} = 2$. The region above this curve violates Bell's inequality. From the plot we see that for temperature up to $19$ mK Bell violation is absent when the transmission efficiency $\eta < 0.35$ ($> 65 \%$ signal loss). For temperature higher than $19$ mK greater value of transmission efficiency is required to observe Bell violation. $B^L_{max}$ falls sharply with increasing signal loss for temperature $T > 21$ mK. For our chosen range of parameters Bell violation is absent when $T \geq 28$ mK.


Though the main purpose of the present work is to show that Bell violation
is indeed possible in the DCE set-up, it is also indeed feasible to conceive
schemes to experimentally measure such Bell violation.
Entanglement in DCE radiation has been quantitatively measured  in a recent experiment in superconducting circuit \cite{broadband_dce}. Experimental studies on DCE radiation show that in our chosen temperature range, the thermal photons are quite challenging to resolve. Nonetheless, quantum correlation is observed for some chosen range of parameters in presence of noise due to thermal photons.
Note that Bell-violating states form a subset of all entangled states
\cite{wiseman}. Hence, a study of entanglement is not equivalent to a study of Bell violation.
In the context of the present set-up,  the range of system parameters for
which Bell violation is obtained is much smaller than that of entanglement.

In the present analysis the Bell operator consists of two measurements $\hat{\Pi}_x \otimes \hat{\Pi}_x$ and $\hat{\Pi}_z \otimes \hat{\Pi}_z$. The measurement $\hat{\Pi}_x$ can be written as $sgn(\hat{q})$ ($\hat{q} \rightarrow$ canonical position operator) \cite{revzen3}.   It is basically the sign of the quadrature and can be measured by homodyne measurement that has been implemented in experiments \cite{jo3, parn}.   The operator $\hat{\Pi}_z$ is the parity operator in spatial basis and it has the same expectation value with the parity operator in the number basis \cite{revzen2}.
 So, it is possible to measure $\langle\hat{\Pi}_z \otimes \hat{\Pi}_z \rangle$ using a number resolving detector. Alternatively, parity can also be measured in the spatial basis using a parity analyser which is the parity sensitive Mach-Zehnder interferometer \cite{space_parity1, space_parity2}.  Schemes for implementing the Mach-Zehnder interferometer  in superconducting coplaner waveguide (CPW) have been proposed \cite{mzi_cqed1}. Detailed discussions
on implementing various components such as mirrors, phase shifts, and photon detectors for coincidence circuits in CPW have been provided (sse, for instance,
\cite{mzi_cqed2}.


Before concluding, it may be pertinent to note the following issue.
There are several important loopholes in the experimental violation of a Bell inequality \cite{larsson}. Two of the
most widely discussed loopholes are the locality loophole and the detection
loophole. While  the locality loophole cannot
be closed in the present set-up, further analysis is required in respect of
the detection loophole here. In general, sufficiently high detector
efficiencies enable closure of the detection loophole in Bell tests involving
parametric down conversion \cite{giustina}. Note that though in case of the set-up 
involving DCE photons that we have considered here, a small magnitude of
Bell violation is obtained in the perturbative regime of the driving amplitude,
such violation still persists under considerable signal loss.
Moreover, our study indicates that Bell violation increases with the driving amplitude, and hence, a detailed study will be required involving a non-perturbative analysis
in conjunction with tolerance to signal loss in order to estimate the
threshold of detector efficiency needed for closure of the detection
loophole.

\section{Conclusions}\label{sec5}

Quantum nonlocality as manifested by the violation of Bell's inequality
represents a basic paradigm of quantum theory, which is of importance for
the test of foundational principles, as well as for potential technological
applications, such as in quantum cryptography. In this work we have studied Bell
 violation by dynamical Casimir photon pairs generated from quantized vacuum by the relativistic motion of a mirror. We have considered the circuit quantum electrodynamical set-up that has been experimentally implemented \cite{jo3}. 
 Though Bell's inequality has been studied earlier theoretically in the noninertial relativistic domain \cite{unruh_bell,cosm_bell1,cosm_bell2},
 experimental verification of such proposals remain beyond the reach of
 current technology. On the other hand, the framework of Bell violation proposed
 in the present work can be probed efficiently in the laboratory \cite{jo1,jo2,jo3,parn}. 
 
The analysis performed in this work is based on the measurement induced spin-like quantum correlations within  Casimir photon pairs. Previously, several theoretical and experimental studies have been performed on Gaussian quantum correlation of DCE photons using homodyne measurement. Our present study focuses on nonlocal quantum correlations between Casimir photon pairs generated through non-Gaussian measurements \cite{revzen2, revzen3}. Such correlations have been shown to be of significance in several domains of quantum information and communication \cite{revzen_telep, pspin_nha, pspin_adesso, singh_bose, treps_nong}. The Bell violation obtained here through the above framework
is thus of direct relevance to several information theoretic protocols.


Let us now briefly summarize the main results of our study. We have analytically derived the expectation value of the Bell operator for DCE radiation, optimized with respect to channel orientations in the context of pseudospin measurements. We have studied the behaviour of Bell violation in terms of experimental parameters such as the driving frequency. We have further considered the effect of local thermal noise in each mode and asymmetry between the entangled modes introduced through the detuning in the frequency of photon pairs. We show that for our chosen parameter range, the violation of Bell's inequality  can be observed up to the temperature about $40$ mK. Our results further show
that the asymmetry between the entangled modes degrades the Bell nonlocality at relatively higher temperature. However, at low temperatures detuning has negligible effect on Bell's inequality violation. 
Finally, we have also derived the expectation value of the Bell operator in presence of signal loss and explored the robustness of Bell nonlocality in this scenario. We show that in the system under consideration, Bell nonlocality is robust up to $65 \%$ signal loss. 

To conclude, in our analysis
we have presented multiple plots showing the variation of Bell's nonlocality with different circuit parameters in presence of local noise, asymmetry between the entangled modes because of nonzero detuning, and signal loss. Our results clearly 
demarcate the parameter regions where Bell nonlocality of Casimir photons can be observed. The choice of parameters considered in the present study is well 
within the perturbative regime. Since Bell violation is seen to rise rapidly
with increase of the driving frequency, it is expected that higher values of the driving parameter would yield significantly larger magnitudes of Bell violation. Our results thus motivate further analysis in the nonperturbative framework. It might be also interesting to consider in future works the
Bell violation in the cQED set-up using other measurement schemes. A comparative
analysis of such studies may lead to an optimal framework for quantum state preparation of Casimir photons, as a vital step towards information processing through the cQED set-up.

{\it Acknowledgements:} ASM acknowledges support from the Project no. DST/ICPS/QuEST/2018/Q-79 of the Department of Science \& Technology, India.

\end{document}